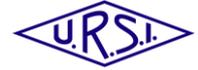

# Type III Radio Bursts from Solar Eruptions and their Connection to GLE and SGRE Events

Nat Gopalswamy[(1)], Anshu Kumari* [(1)], and Pertti A. Mäkelä [(1,2)]
(1) NASA Goddard Space Flight Center, Greenbelt, Maryland, USA; e-mail: nat.gopalswamy@nasa.gov
(2) The Catholic University of America, Washington DC, USA

## Abstract

We report on the close similarity of coronal mass ejection (CME) properties in ground level enhancement (GLE) in solar energetic particle (SEP) events and sustained gamma-ray emission (SGRE) from the Sun as indicated by low frequency type III radio bursts observed in the interplanetary medium. The complex type III bursts have an average 1 MHz duration of 36 and 34 min in the SGRE and GLE events, respectively. Similarly, the CMEs underlying SGRE and GLE have average space speeds of 1866 and 2084 km/s, respectively. These are larger than the corresponding values (32 min, 1407 km/s) for a control sample of type III bursts associated with frontside halo CMEs with sky-plane speed >800 km/s. These results are consistent with the idea that energetic CME-drive shocks accelerate particles to very high energies that are responsible for GLE and SGRE events.

## 1 Introduction

Type III radio bursts are produced in the corona and interplanetary medium when nonthermal electrons accelerated near the Sun propagate along open magnetic field lines causing coherent plasma emission. The source of nonthermal electrons distinguishes two varieties of type III bursts: (i) eruption type III bursts due to electrons accelerated in the flare site, and (ii) storm type III bursts caused by nonthermal electrons accelerated in the interchange reconnection region between active region field lines and open field lines present adjacent to active regions. Here we focus on the eruption type III bursts, which are good indicators of the early phase of solar eruptions. In particular we consider low-frequency type III bursts observed in space at frequencies below the ionospheric cutoff (~15 MHz). These bursts are referred to as complex type III bursts (Reiner et al. 2000) [1] or type III-l bursts (Cane et al. 2002) [2] to reflect the long-duration (typically>15 min) and low-frequency (<14 MHz) nature of these bursts. These bursts appear as prominent features in the dynamic spectra obtained by spaceborne radio telescopes such as the radio and plasma wave experiment (WAVES) on board Wind [3] and STEREO [4] spacecraft.

Complex type III bursts are thought to be indicative of large solar energetic particle (SEP) events [2,5] with the implication that the SEPs may be accelerated at the flare site similar to the electrons causing the type III bursts. A closer examination of type III bursts associated with SEP events revealed that one cannot discriminate between the flare and shock paradigms for SEP events based on complex type III bursts. Furthermore, the occurrence of a type III burst does not seem to be a sufficient condition for a SEP event [6]. Gopalswamy et al. [7] investigated the relation between type III bursts and ground level enhancement (GLE) events. They found that the eruption type III bursts associated with GLEs are all complex, of long-duration, and intense at low-frequencies. Since the presence of type III bursts is not a sufficient condition for the occurrence of a SEP event, the association is taken to simply reflect the fact that GLEs are associated with large flares that result in complex type III bursts, but the production of high-energy particles themselves may not be at the flare site [8]. GLE events indicate the acceleration of protons to GeV energies. One other class of energetic events from the Sun involving high-energy protons (>300 MeV) is the sustained gamma-ray emission (SGRE) from solar eruptions. It has been shown that GLE events and SGRE events have very similar coronal mass ejection (CME) properties [9-10], indicating that the CME-driven shocks accelerate particles to high energies that cause the two phenomena. GLE events are very rare, which is more so during the weak solar cycle 24: only two GLEs were observed in this cycle. One of the reasons for the lack of GLE events is the poor magnetic connectivity of the eruption to an Earth observer [11]. On the other hand, SGRE photons are not affected by magnetic connectivity and therefore a lot more SGRE events are observed in solar cycle 24 than the number of GLE events. The question is whether the type III bursts in SGRE-related eruptions share properties with GLE events. In order to check this we consider type III bursts associated with three sets of events: (i) GLE events, (ii) SGRE events, and (iii) frontside halo CMEs with speed $\geq$ 800 km/s associated with type II bursts.

## 2 Data Selection

Data on 18 GLEs has already been published in cycles 23 [7] and 24 (Gopalswamy et al. 2013;2018). To this list we add the lone GLE event that occurred on 28 October 2021, bringing the sample size to 19 for a better statistics. The SGRE events are observed by the Large Area Telescope (LAT) [12] on board the Fermi satellite starting in 2008. More than 30 SGRE events have been cataloged [13-14]. Of these we use 24 SGRE events that have duration $\geq$3 hours [10,11]. We compile the properties of type III bursts and CMEs associated with the GLE and SGRE events. As a control sample we consider type III bursts associated with frontside halo CMEs with speeds >800 km/s and associated with low frequency type II radio bursts listed in (https://cdaw.gsfc.nasa.gov/CME_list/radio/waves_type2.html). The reason for requiring type II burst association is to make sure that the frontside halos are driving shocks as in the case of the GLE and SGRE events.

Figure 1 shows an SGRE event, which shows various phenomena associated with the 2014 February 25 eruption: a complex type III burst, a type II burst, an intense SEP event, and an X4.9 large soft X-ray (SXR) flare. The eruption occurred from the east limb (S12E82) and was observed by SOHO and STEREO instruments. Note that type III burst ends before the SXR peak. On the other hand, the SGRE event peaks well after the SXR peak. The duration of the type II burst is very similar to that of the SGRE event. The eruption resulted in an ultrafast CME (2147 km/s in the sky plane), an SGRE event, and a large SEP event. The type III burst starts at 00:48 UT on 2014 February 25 and ends at 01:25 UT, thus having a duration of ~37 min. In a similar way, we measure the duration of the type III bursts in all three sets of events and compile the sky-plane speeds of the underlying CMEs from the CDAW CME catalog (https://cdaw.gsfc.nasa.gov). We compare the type III and CME properties among the three sets.

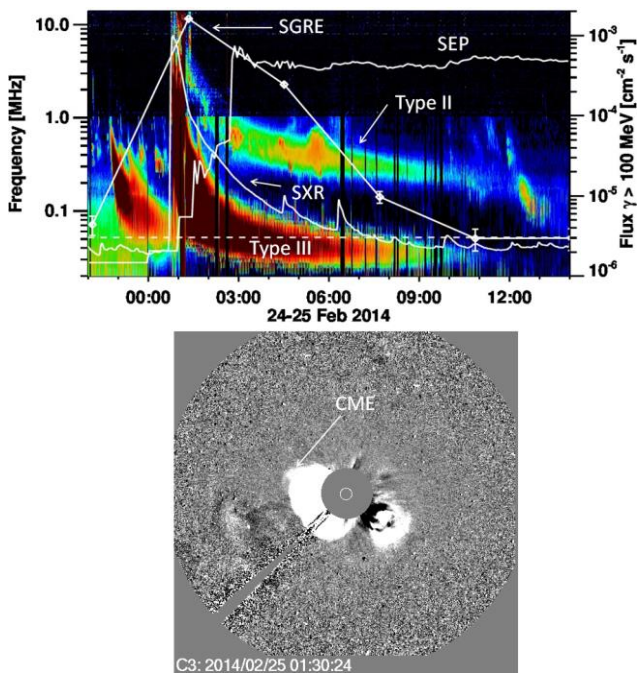

**Figure 1.** (top) A Wind/WAVES radio dynamic spectrum obtained during 2014 February 24-25 showing the type II and complex type III bursts. The curve with data points denoted the flux of gamma-rays at energies >100 MeV as detected by Femi/LAT (right side scale). GOES proton intensity curve (marked SEP) and the soft X-ray (SXR) light curve are shown in arbitrary units. (bottom) A snapshot of the underlying CME observed by the SOHO/LASCO outer coronagraph C3.

## 3 Results

Figure 2 compares the durations of type III bursts associated with SGRE and GLE events. The durations are very similar, confirming the close relation between the SGRE and GLE events in that both involve eruptions resulting in high energy particles. Figure 3 shows the durations of type III bursts associated with frontside halo CMEs. Clearly, the durations are much smaller at 14 MHz compared to SGRE and GLE type III bursts. The 1 MHz duration is only slightly smaller.

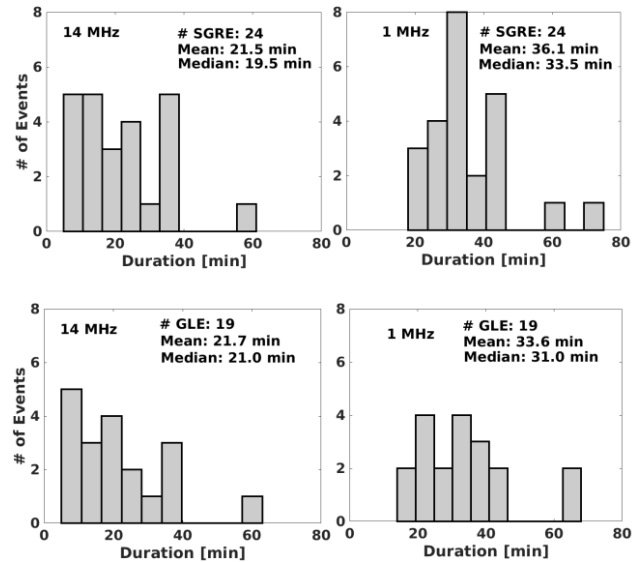

**Figure 2.** Distributions of type III burst durations for SGRE events (top row) and the GLE (bottom row) events at 14 MHz (left columns) and 1 MHz (right columns). The mean and median values are also shown on the plots.

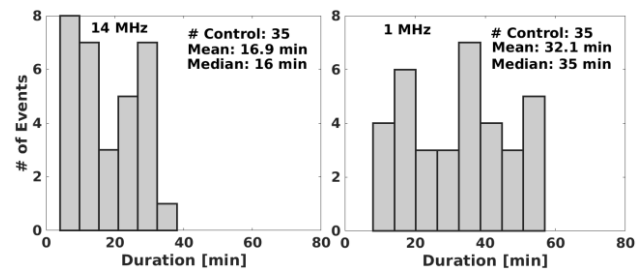

**Figure 3.** Distributions of type III burst durations for the frontside halo CMEs at 14 MHz (left) and 1 MHz (right). The mean and median values are also shown on the plots.

We also compare the sky-plane speeds of the three populations in Figure 4 (top panel). We see that the average speeds of CMEs associated with SGRE (~1700 km/s) and GLE events (~1900 km/s) are very similar and much higher than the speed (~1250 km/s) of the frontside halo CMEs. To get a better comparison, we have shown the three-dimensional speeds of CMEs in the bottom panel of Figure 4. These speeds were obtained either using the cone model [15] or by fitting a graduated cylindrical shell (GCS) flux rope to the SOHO and STEREO coronagraph images [16].

As expected, the GLE CMEs have the highest space speed (~2000 km/s) only slightly higher than the SGRE CME speed. On other hand, both SGRE and GLE CMEs have speeds much higher than that of the frontside halos. The very high speeds of the SGRE and GLE CMEs is consistent with the fact that they drive strong shocks that accelerate particles to the highest energies.

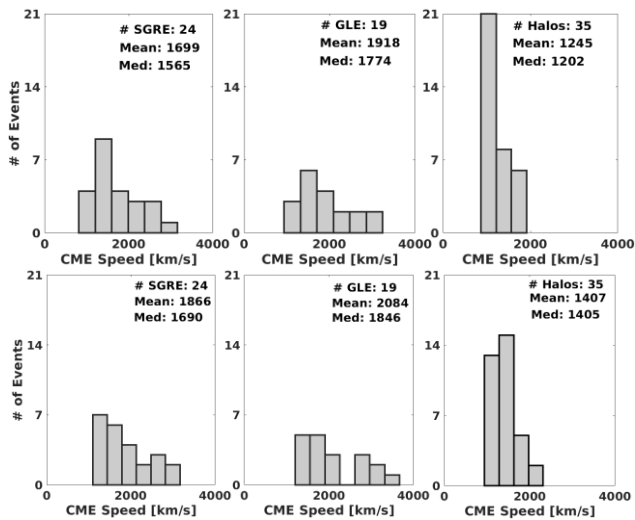

**Figure 4.** (top) Distributions of sky-plane speeds of frontside halo CMEs (left), CMEs associated with SGREs (middle), and CMEs associated with GLE events. (bottom) The corresponding space speeds obtained from a cone model or using GCS fit to SOHO and STEREO observations. The mean and median speeds (km/s) are indicated on the plots.

## 4 Summary and Conclusions

We examined the properties of eruptive phenomena such as complex type III bursts and CMEs associated with SGRE and GLE events. We used 24 Fermi/LAT SGRE events that had durations ≥ 3 hours and 19 GLE events (16 from solar cycle 23, two from cycle 24, and one from the rise phase of cycle 25). We also considered a control sample of frontside halo CMEs that are shock-driving (associated with type II radio bursts). The type III burst durations and CME speeds are nearly identical between SGRE and GLE events, confirming that the underlying CMEs are the most energetic and accelerate particles to very high energies. The small number of GLE events in solar cycle 24 compared to the 24 SGRE events is a result of the magnetic connectivity required for the GLE events. There is no such connectivity requirement for SGRE photons, so energetic eruptions that are not well connected to an Earth observer can produce SGRE events, even though the associated high-energy particles do not reach an Earth observer. The control sample also represents an energetic population because the underlying CMEs are halos and shock-driving. The 1 MHz duration of type III bursts associated with the control sample is only slightly smaller than those of SGRE and GLE events. However, the average CME speed is significantly lower than those of SGRE and GLE events. The long duration type III bursts are indicative of the large eruptions but may not indicate the acceleration of >300 MeV particles. On the other hand, the CME speed is able to discriminate eruptions that result in high energy particles responsible for SGRE and GLE events.

## Acknowledgements

We benefited from the open data policy of SOHO, STEREO, GOES, and Wind teams. Work supported by NASA's STEREO project and the Living With a Star program. PM is supported in part by the NSF grant AGS-2043131. We thank Sindhuja Gunaseelan for checking the GCS speeds.

## References

[1] M. J. Reiner et al. *ApJ* **530** 1049, 2000, doi: 10.1086/308394
[2] H. V. Cane, W. C. Erickson, N. P. Prestage, *JGR*, **107** Issue A10, CiteID 1315, 2002, doi: 10.1029/2001JA000320
[3] J.-L. Bougeret et al., *Space Sci. Rev.*, **71** 231, 1995, doi: 10.1007/BF00751331
[4] J.-L. Bougeret et al., *Space Sci. Rev.* **136** 487, 2008, doi:10.1007/s11214-007-9298-8.
[5] R. J. MacDowall, et al., *IAUS* **257** 335, 2009, doi: 10.1017/S1743921309029512
[6] N. Gopalswamy, P. Mäkelä, *ApJ* **721** L62, 2010, doi: 10.1088/2041-8205/721/1/L62
[7] N. Gopalswamy, H. Xie, S. Yashiro, S. Akiyama, P. Mäkelä & I. G. Usoskin, *Space Sci. Rev.* **171** 23, 2012, doi: 10.1007/s11214-012-9890-4
[8] E.W. Cliver, A. and G. Ling, ApJ **690** 598, 2009, doi: 10.1088/0004-637X/690/1/598
[9] N. Gopalswamy, Mäkelä P., Yashiro S. et al., *ApJL* **868** L19, 2018, doi: 10.3847/2041-8213/aaef36
[10] N. Gopalswamy, P. Mäkelä, S. Yashiro et al., *JPhCS* **1332** 012004, 2019, doi: 10.1088/1742-6596/1332/1/012004
[11] N. Gopalswamy, S. Yashiro, P. Mäkelä, H. Xie, and S. Akiyama, *ApJ* **915**, 82, 2021, doi: 10.3847/1538-4357/ac004f
[12] M. Ajello, A. Albert A. Allafort. et al. *ApJ* **789** 20, 2014, doi: 10.1088/0004-637X/789/1/20
[13] G. H. Share, R. J. Murphy, A. K. Tolbert et al. *ApJ* **869** 182, 2018, doi: 10.3847/1538-4357/aaebf7
[14] M. Ajello et al.. *ApJ* **252** 13, 2021, doi: 10.3847/1538-4365/abd32e
[15] H. Xie, Ofman, L., and Lawrence, G., *JGR* **109**, A03109, 2004, doi: 10.1029/2003JA010226
[16] A. Thernisien, *ApJS* **194**, Issue 2, article id. 33, 2011, doi: 10.1088/0067-0049/194/2/33